\documentclass[useAMS,usenatbib]{mn2e}
\usepackage{graphicx}
\title[Heterogeneity of solid neutron-star matter]{Heterogeneity of solid
neutron-star matter: transport coefficients and neutrino emissivity}
\author[P. B. Jones]{P. B. Jones\thanks{E-mail: p.jones1@physics.ox.ac.uk} \\
Department of Physics, Denys Wilkinson Building, University of Oxford,
Keble Road, Oxford, OX1 3RH, UK}

\begin{document}

\date{ }

\pagerange{\pageref{firstpage}--\pageref{lastpage}} \pubyear{2002}

\maketitle

\label{firstpage}

\begin{abstract}
Calculations of weak-interaction transition rates and of nuclear formation
enthalpies show that in isolated neutron stars, the solid phase, above
the neutron-drip
threshold, is amorphous and heterogeneous in nuclear charge.  The neutrino
emissivities obtained are very dependent on the effects of proton shell
structure but may be several orders of magnitude larger than the electron
bremsstrahlung neutrino-pair emissivity at temperatures $\sim 10^{9}$ K.
In this phase, electrical and thermal conductivities are much smaller than
for a homogeneous {\it bcc} lattice.
In particular, the reduced electrical conductivity, which is also
temperature-independent, must have significant consequences for the evolution
of high-multipole magnetic fields in neutron stars.
\end{abstract}


\section{Introduction}

The very extensive pulsar and X-ray source observations now being made
require, for their interpretation, an understanding of the condensed-matter
physics of neutron-star solid phases.  The radiative opacity of a very thin
surface layer of depth $\sim 10^{3}$ cm and matter density
$\rho \stackrel{<}{\sim} 10^{7}$
g cm$^{-3}$ largely determines the temperature difference between the surface
and the interior of the star (see Potekhin, Chabrier \& Yakovlev 1997, also
Potekhin \& Yakovlev 2001, for recent calculations and reviews of earlier
work). Atoms in the higher-density part of this layer are completely
ionized but the electron Fermi momentum is less than or of
the order of $1$ MeV/c.  The whole layer contains no more than $10^{-10}
\rm{M}_{\odot}$ and has physical properties which are important only with respect
to its radiative opacity and, possibly, in connexion with the composition of
the neutron-star
atmosphere.  In the next layer, with matter densities up to the neutron-drip
threshold $\rho_{nd} = 4.3 \times 10^{11}$ g cm$^{-3}$, depth $\sim 10^{4}$
cm and mass $\sim 10^{-5} \rm{M}_{\odot}$, the electrons form a relativistic
Fermi gas
whose transport coefficients (electrical and thermal conductivities $\sigma$ and
$\kappa$) are dependent on the nuclear composition and degree of order of the
solid.  This is also true of the neutron-drip region with densities above
$\rho_{nd}$
which occupies most of the crust volume (depth $\sim 10^{5}$ cm and mass
$\sim 10^{-2} \rm{M}_{\odot}$).  The fraction of the stellar volume
concerned here is so large that its electrical conductivity must be relevant to
the evolution of high-multipole components of the magnetic field.
Pinning of superfluid neutron
vortices by nuclei in this region is believed to be the origin of pulsar glitch
phenomena (Anderson \& Itoh 1975; Ruderman 1976).  The neutron-drip region
contributes almost all the mechanical rigidity of the crust and its
failure under Maxwell or other stresses is assumed to be involved in a
number of X-ray emission phenomena, for example, the soft gamma repeaters
(SGR; Thompson \& Duncan 1995, 1996) and the persistent emission of the
anomalous X-ray pulsars (AXP; Thompson et al 2000).

The canonical picture of the solid, both above and below $\rho_{nd}$, in a
neutron star that has not been subject to a long period of accretion since
formation,
is of a homogeneous {\it bcc} lattice of even-$Z$ nuclei, locally in
complete weak-interaction equilibrium.  Certainly above,
and possibly immediately below, $\rho_{nd}$, the nuclei are those with
closed proton shells. The equation of state below $\rho_{nd}$ has
been derived by extrapolations of nuclear parameters from experimentally
accessible regions of neutron excess (see Haensel \& Pichon 1994 who also 
summarize earlier work).  Pethick \& Ravenhall (1995) give a general review of
solid phase properties and observe that, above $\rho_{nd}$, there have been
two distinct approaches to  the problem of deriving the equation of state.
Microscopic calculations of the single-particle states for neutrons and protons
inside a Wigner-Seitz cell were described in the classic paper of Negele
\& Vautherin (1973) and give a complete description of the system apart
from the superfluid energy gap.  But this approach has not been followed by
later workers who have adopted a compressible liquid-drop model (CLDM)
with various Skyrme pseudopotentials (Lattimer et al 1985, Douchin \&
Haensel 2001).  Although the CLDM nuclear charge is a continuous variable,
the existence of shell effects and of proton pairing require that the crust
be composed of successive homogeneous layers of even-$Z$ nuclei.
In fact, weak-interaction
equilibrium cannot be exact owing to the rapid decrease of weak transition
rates as the star cools, which is caused by the potential barrier present in 
those transitions
to odd-$Z$ nuclei.  Flowers \& Ruderman (1977) noted that, in consequence,
at least a small fractional concentration of nuclei with charge deviation
$\Delta Z = \pm 2$ from the homogeneous lattice must be present 
in metastable equilibrium as point defects.  But it has been
usual to assume that the value of the impurity parameter,
\begin{eqnarray}
Q=\sum_{i}a_{i}(Z_{i}-\bar{Z})^2,
\end{eqnarray}
defined for a distribution of nuclear charges $Z_{i}$ with fractional
concentrations $a_{i}$ and mean $\bar{Z}$,
is, in most contexts, negligibly small (several orders of magnitude less than
unity).

The extent of heterogeneity in $Z$ at densities $\rho < \rho_{nd}$ has been
investigated by
several authors.  Jog \& Smith (1982) and De Blasio (2000) have examined the
structure of the interface
between successive homogeneous layers.  The nature of the constraints under
which equilibrium is defined was considered by Jones (1988) who concluded
that a distribution in $Z$ must be a feature of the state reached as the
star cools.  More recently, calculations of the point defect concentration
have been made by De Blasio \& Lazzari (1998).

Formation enthalpies were calculated by Jones (1999a, 2001)
for a number of point-defects in solid neutron-star matter at
densities $\rho > \rho_{nd}$.  The enthalpies obtained were small and it was
argued that an amorphous
heterogeneous solid phase must be formed and should persist as the star cools,  
the melting transition being replaced by a glass transition temperature region.

The conclusion that heterogeneity in $Z$ exists both below and above
$\rho_{nd}$ does not seem to have been widely accepted.  Undoubtedly,
the assumption of a {\it bcc} lattice, homogeneous in $Z$, is attractive
because it can be simply
stated and provides a clear basis for calculations, such as those of
transport coefficients.  It is also the case that arguments about the
role of proton
shell-structure and the approach to weak-interaction equilibrium were
made only qualitatively by Jones (2001, hereafter Paper I) and were not
supported by detailed
calculation.  Nonetheless, even though amorphous heterogeneous structures
are unattractive owing to their greater complexity, if they represent physical
reality, it is necessary to consider their effect on the stress-response
of the solid and to
define as well as possible the consequent degree of uncertainty in
calculations of transport coefficients and neutrino emissivities.  The
present paper contains the results of those detailed calculations which
were absent from the previous papers.  In Section 2, we consider how the
effects of proton shell-structure can be included, quantitatively, in the
compressible liquid-drop model of nuclei by the Strutinski procedure
(see Ring \& Schuck 1980) to obtain formation enthalpies for
nuclei in the interval $20 \leq Z \leq 50$.  Section 3 gives estimates
of the weak-interaction transition rates between these nuclei and, from
an initial temperature of $5 \times 10^{9}$ K, which is in the vicinity of
the melting
temperature or glass transition region of the system, describes how the
$Z$-distribution evolves with temperature and time.  It also gives estimates
of the very broad range of possible neutrino emissivities associated with
these transitions which have so far been neglected in all published
calculations of neutron-star cooling.

This paper is concerned with the nature of the crust in isolated neutron
stars which lack the long period of mass-accretion of binary systems.  Factors
such
as accretion through fall-back at formation are ignored.  Therefore, it does
not consider
neutron stars in those binary systems where the rate of mass transfer is
large enough to replace the whole crust, below and above $\rho_{nd}$
(see Schatz et al 1999).  The extent to which the results obtained here may
be relevant to such systems is discussed briefly in Section 6.
Under the physical conditions considered in the present paper,
pycnonuclear reactions were examined previously (Jones 2002) and were found
to have negligible transition rates in the solid phase.  The reason for this
is that the intermediate state formed in a solid by fusion of nuclei with
charges $Z_{1}$ and $Z_{2}$ consists
of a monovacancy and a point-defect of charge $Z_{1}+Z_{2}$.  At the highest
matter density, $8.8\times 10^{13}$ g cm$^{-3}$, assumed in Sections 2 \& 3,
for example, this state has a 17 MeV formation enthalpy.  (This assumes that
analogues of the standard lattice point-defects exist in amorphous solids,
though they may be short-lived at high temperatures.)
States with even
higher formation enthalpy would result from successive fusion reactions,
indicating that processes such as pinning-induced nuclear rod formation
(Mochizuki, Oyamatsu \& Izuyama 1997) are not significant.  However, the
direct formation of lower-dimensional nuclear structures at temperatures
of the order of $10^{10}$ K, 
in a density interval between the spherical nuclear phase
and the continuous liquid core of the star, is predicted for many models
of nuclear matter (Lorenz, Ravenhall \& Pethick 1993; Oyamatsu 1993;
but see also Douchin \& Haensel 2000).
Lorenz et al noted that the geometrical form of these structures would
allow weak-interaction transitions but gave no estimate of the neutrino
emissivity.  For completeness, a brief calculation of this emissivity is
given in Section 4.  It is less significant than that associated with
the region of spherical nuclei.

The density region below $\rho_{nd}$ is reconsidered in Section 5 using
the binding energy compilation of M\o ller, Nix \& Kratz (1997), but with
inconclusive results for values of $Q$.  It is possible to state only that
values $Q\approx 1$ are probable, with $Q\gg 1$ in limited regions.  This
is not too serious a problem for most neutron star calculations owing to
the limited depth of the region below $\rho_{nd}$.   Values of $Q$ computed
for $\rho > \rho_{nd}$, the region
which occupies most of the crust volume, are given in Table 2 and their
significance is considered in Section 6.

\section[]{Shell effects in the distribution of formation enthalpies}

Formation enthalpies for point-defect structures of impurity nuclei with
charge $Z_{i}$ in an otherwise homogeneous {\it bcc} lattice of charge
$Z$, were obtained in previous work (Paper I) at densities above $\rho_{nd}$.
The method of
calculation followed that used earlier for monovacancies (Jones 1999b),
the essence of which was application of the Feynman-Hellmann theorem
(Slater 1963) to find the lattice displacements in the vicinity of the defect.
Nuclei were described by the compressible liquid-drop model (CLDM)
of Lattimer et al (1985) with these authors' Skyrme pseudo-potential for
bulk nuclear matter and their expression for the thermodynamic potential
per unit area of nuclear surface.   Lattice-site displacements in the vicinity
of an impurity are determined, principally, by properties of the high-density
relativistic electrons at $\rho > \rho_{nd}$.  
The Coulomb-electron stress-tensor has isotropic components which are
between one and two orders of magnitude larger than the off-diagonal.
Also, the inverse of the electron-screening wavenumber is larger than the
{\it bcc} lattice constant.  Consequently, lattice-site displacements are
such that the electron density, averaged over a volume of the order of the
Wigner-Seitz cell, adjusts to values almost exactly equal to the mean electron
density of
the undisturbed lattice.  We refer to Paper I for more complete discussions
of these and other features of the formation enthalpy calculations.  Our
primary assumption about the distribution of formation enthalpies is based
on these considerations.  It is that the formation enthalpy for a nucleus
of charge $Z_{i}$ in an amorphous heterogeneous solid of mean charge
$\bar{Z}$, or in a liquid with the same nuclear charge distribution, is
satisfactorily approximated by that calculated
for an impurity nucleus of charge $Z_{i}$ in a homogeneous
lattice of charge $\bar{Z}$.

The formation enthalpies were given in Paper I with reference to that for the
homogeneous lattice charge.  With the exclusion of integral multiples of
the neutron and electron chemical potentials, they were expressed as
$H_{FZ} = C(Z - \bar{Z})^{2}$ for charge $Z$.
Values of the constant $C$ are given here in Table 1 at several matter
densities with, for convenience, the parameters of the lattices concerned.
The values of $\rho$ chosen exclude the region immediately above $\rho_{nd}$
because it represents a relatively small interval of depth in the solid crust.
The CLDM parameters used by Lattimer et al fit the ground states of
laboratory even-even nuclei and so include the effects of pairing
interactions. Thus our expression for $H_{FZ}$ neglects shell
effects, including the unpaired proton in odd-$Z$ nuclei.
The investigation by Negele \& Vautherin (1973)
revealed a proton shell structure very similar to that of laboratory
nuclei.  Although shell energy-differences are modified in the neutron
continuum, the shell ordering (see Figure 5 of their paper) is changed only
in that $1\rm{d}_{3/2}$ and $1\rm{f}_{5/2}$ respectively, precede
$2\rm{s}_{1/2}$ and $2\rm{p}_{3/2}$.  We are unaware of any published
sequence of single-particle energy levels for nuclei beyond the neutron-drip
threshold apart from those contained in that paper.  However, Chabanat et al
(1998) have shown by calculation of two-neutron separation energies
that shell effects at neutron numbers $N=50, 82$ remain significant
immediately below the neutron-drip threshold.
There is no doubt that our expression
for $H_{FZ}$ ought to be modified by shell structure, but the
form of these changes was considered only qualitatively in Paper I.

At matter densities $\rho > \rho_{nd}$, there is a neutron continuum with
chemical potential $\mu_{n} > 0$.  It is degenerate, except for a small
volume with $\rho \approx \rho_{nd}$, and is superfluid at temperatures
$0 < T < T_{cn}^{e}$.  Nuclei can be
most simply viewed as bound states of protons embedded in this system,
charge-neutralized by an almost uniform relativistic electron gas.
The formation enthalpy differences $H_{FZ+1} - H_{FZ}$ can,
in principle, be affected by neutron shell-structure because the change in
$Z$ will be associated with a change in nuclear radius $r_{N}$ and, possibly,
a change in the number of neutron single-particle states at negative energy.
But we shall assume that neutron transitions from the continuum just above the
zero of energy to states just below do not contribute to discontinuities
in the formation enthalpy differences with which this paper is concerned.
Thus we consider the effects of proton shell-structure only, and regard
the neutrons within the nuclear volume as merely a part of the superfluid
continuum, though with increased density and a locally modified superfluid
energy gap.

\begin{table*}
\centering
\begin{minipage}{140mm}
\caption{Properties of CLDM lattice nuclei in equilibrium with a uniform neutron
liquid of number density $n_{n}^{e}$.  The matter density is $\rho$, and
$\bar{Z}$ is here the uniform nuclear charge, a CLDM continuous variable.
The Fermi wavenumbers are
$p_{Fn}$ for neutrons within the nuclear volume and $p_{Fe}$ for the electrons.
The radii of the nucleus and Wigner-Seitz cell are $r_{N}$ and $r_{WS}$,
respectively.  The lattice Debye and melting temperatures are $T_{D}$ and
$T_{m}$. The energy gap of the neutron continuum is $\Delta_{n}^{e}$, and
$C$ is the formation enthalpy constant.}
\begin{tabular}{@{}lllllllllll@{}}
\hline
$n_{n}^{e}$ & $\rho$ & $\bar{Z}$ & $p_{Fn}$ & $p_{Fe}$ & $r_{N}$ & $r_{WS}$ &
$T_{D}$ & $k_{B}T_{m}$ & $\Delta_{n}^{e}$ & $C$\\
$10^{-3}$ fm$^{-3}$ & $10^{13}$ g cm$^{-3}$ & & fm$^{-1}$ & fm$^{-1}$ & fm &
fm & $10^{9}$ K & MeV & MeV & MeV\\
\hline
7.8 & 1.6 & 34.65 & 1.47 & 0.231 & 5.82 & 27.1 & 1.8 & 0.36 & 0.77 & 0.0142\\
18.4 & 3.7 & 35.13 & 1.48 & 0.286 & 6.28 & 22.0 & 2.3 & 0.46 & 1.10 & 0.0096\\
43.6 & 8.8 & 34.26 & 1.49 & 0.363 & 7.10 & 17.15 & 2.9 & 0.56 & 0.67 & 0.0051\\
\hline
\end{tabular}
\end{minipage}
\end{table*}

Shell effects in the formation enthalpies of nuclei with $20 \leq Z \leq 50$
are estimated here using the Strutinski procedure, following fairly closely
the account given by Ring \& Schuck (1980).  If a purely notional
single-particle level sequence were to be associated with the CLDM
approximation, its level density would be a monotonic and smoothly
varying function of $Z$ or of single-particle energy.  Real
single-particle level densities are not like this.  Thus the procedure
starts from a computed level density $g(\epsilon)$ and single-particle
energy sum $E_{sp}$ and generates an averaged level density
$\tilde{g}(\epsilon)$ and energy sum $\tilde{E}_{sp}$.
Hence the formation enthalpy
deviation from its CLDM value at a specific $Z$ is given by
\begin{eqnarray}
H_{FZ} = (H_{FZ})_{\rm{CLDM}} + E_{sp} - \tilde{E}_{sp} + \epsilon_{pq}.
\end{eqnarray}
The average of this expression, for even-$Z$ nuclei, over an interval of
$Z$ would then be
$(H_{FZ})_{CLDM}$, the underlying CLDM formation enthalpy.
The correction terms in this expression should, of course, be enthalpies
because the weak-interaction transitions occur at constant pressure. But
the error in replacing enthalpies by energies here is not large and is
certainly less serious than that inherent in the neglect of configuration
mixing, to which we shall refer later.  The first three terms in the
right-hand side of equation (2) give the shell-corrected formation enthalpy
for even-$Z$ nuclei.  The remaining term, nonzero only for odd-$Z$ nuclei,
is the excitation energy of a single proton quasiparticle.
The shell-model energy is
\begin{eqnarray}
E_{sp} = \int^{\lambda}_{-\infty}\epsilon g(\epsilon)d\epsilon,
\end{eqnarray}
where $\lambda$ is the proton chemical potential and
\begin{eqnarray}
g(\epsilon) = \sum_{i}\delta(\epsilon - \epsilon_{i})
\end{eqnarray}
is the density of single-particle states.  The averaged shell-model energy
is given by the Strutinsky-averaged density of states $\tilde{g}$ defined by
the averaging procedure
\begin{eqnarray}
\tilde{g}(\epsilon) = \int^{\infty}_{-\infty}g(\epsilon^{\prime})
f\left(\frac{\epsilon-\epsilon^{\prime}}{\gamma}\right)
d\left(\frac{\epsilon-\epsilon^{\prime}}{\gamma}\right).
\end{eqnarray}
It is
\begin{eqnarray}
\tilde{E}_{sp} = \int^{\tilde{\lambda}}_{-\infty}\epsilon \tilde{g}(\epsilon)
d\epsilon,
\end{eqnarray}
where the modified chemical potential $\tilde{\lambda}$ remains to be
determined.  It is clearly necessary that successive applications of the
averaging procedure should leave $\tilde{g}$ unchanged.  A suitable
class of functions satisfying this condition is formed by products
of gaussian functions with generalized Laguerre polynomials.  We assume
a specific order of polynomial,
\begin{eqnarray}
f(x) = \frac{1}{\sqrt{\pi}}\rm{e}^{-x^{2}}\left(\frac{15}{8}
- \frac{5}{2}x^{2} + \frac{1}{2}x^{4}\right),
\end{eqnarray}
and refer to Ring \& Schuck for further details and for a tabulation of them.
An averaged occupation number $\tilde{n}_{i}$ is defined for
each proton state,
\begin{eqnarray}
\tilde{n}_{i} = \int^{\frac{\tilde{\lambda}-\epsilon_{i}}{\gamma}}_{-\infty}
f(x)dx
\end{eqnarray}
and the constraint
\begin{eqnarray}
\sum_{i}\tilde{n}_{i} = Z
\end{eqnarray}
is used to determine the modified chemical potential $\tilde{\lambda}$.  The
parameter $\gamma$ defines the width of the averaging function. It is
chosen to satisfy, so far as possible, the condition that the averaged energy
$\tilde{E}_{sp}(\gamma)$ should be independent of it.

The scheme requires a set of proton single-particle levels $\epsilon_{i}$.
As in Paper I, and in order to make use of the values of
the parameter $C$ calculated in that work, we obtain  approximate sets
of homogeneous {\it bcc} lattice
parameters by using the CLDM approximation, closely following Lattimer et al
(1985).  These are given in Table 1 and form a basis for our calculation
of formation enthalpy differences.  For each  matter density, we obtain a
set of $\epsilon_{i}$ from the work of Negele \& Vautherin (1973; fig. 5).
Values of the width $\gamma$ in the interval $2 \leq \gamma \leq 6$ MeV have
been investigated.  The choice is to some degree a matter of compromise
because the stationarity condition is not perfectly satisfied by
$\tilde{E}_{sp}(\gamma)$ at a common $\gamma$ for all $Z$ in the
range $20 \leq Z \leq 50$.  Thus we
have adopted a fixed value $\gamma = 4$ MeV for all sets of $\epsilon_{i}$.
The first three terms in the right-hand side of equation (2) are then
calculated
from the values of the constant $C$ given in Table 1 and from the procedure
described above.  The remaining term $\epsilon_{pq}$ presents some difficulty.
It is recognized that its value depends on the extent to which the shell
concerned is filled but for simplicity, we have assumed a constant value
for all odd-$Z$ nuclei, with $\epsilon_{pq} = 0$ for even-$Z$ nuclei.

A typical set of $H_{FZ}$ values given by equation (2) is shown in Fig. 1,
which is for a matter
density $\rho = 3.7 \times 10^{13}$ g cm$^{-3}$.  The enthalpies exclude
integral multiples of the electron and neutron chemical potentials. There
are pronounced minima in formation enthalpy at the closed-shell charges
$Z = 20, 34, 40, 50$.  These minima are also present at the other densities
listed in Table 1 though with rather different relative values of formation
enthalpy.  For the
formation of a thermal equilibrium population of nuclei, or for the
calculation of weak-interaction transition rates, we require only the
differences $H_{FZ+1} - H_{FZ}$.  Also shown are
$(H_{FZ})_{CLDM}$ values corrected only by the $\epsilon_{pq}$ term.
Comparison of these two sets confirms that, as suggested in Paper I,
shell effects are a more important source of formation enthalpy differences
between nuclei of neighbouring $Z$ than the CLDM bulk nuclear matter term.

\begin{figure}
\scalebox{0.47}{\includegraphics{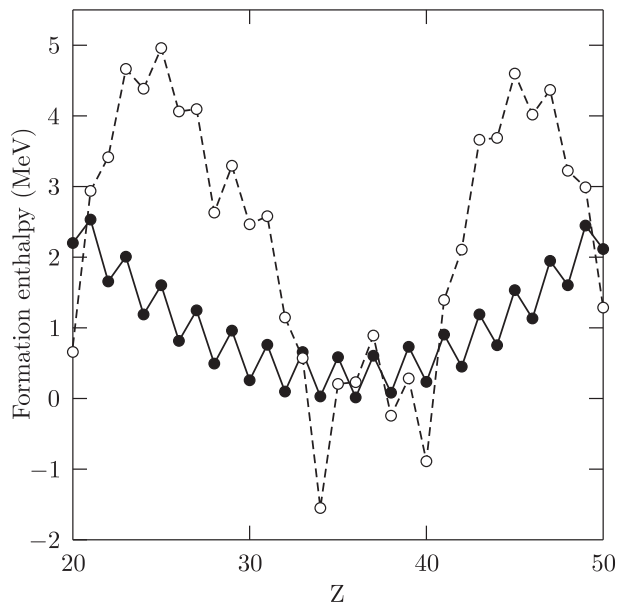}}
\caption{Formation enthalpies $H_{FZ}$ are given for nuclei $20\leq Z \leq 50$
at matter density $\rho = 3.7 \times 10^{13}$ g cm$^{-3}$.  Those represented
by open circles have been obtained using the Strutinsky procedure whereas
the solid circles are the $(H_{FZ})_{CLDM}$ values corrected only for the
unpaired proton.  In both cases, this term is $\epsilon_{pq}=0.6$ MeV.
In the former case, proton shell closures at $Z = 20, 34, 40, 50$ are
associated with very marked enthalpy minima.}
\end{figure}

The fact that the important closed shells are at $Z=34, 40$ is a
consequence of our choice, in Paper I, of the Lattimer et al CLDM
which produces $\bar{Z}\approx 35$ in the density interval considered here
(see also Pethick \& Ravenhall 1995, Fig. 2).  If, instead, the Douchin
\& Haensel CLDM which gives typically $\bar{Z}\approx 45$ had been used,
we anticipate that the general form of Fig.1 would be unchanged except
for a displacement to higher $Z$, with the important closed shells
being those at $Z=40, 50$.
The shell effects of this paper have been obtained using a specific
scheme of calculation
and it is worth considering the extent to which they are physically
realistic.  The values of $\rho$ concerned are subnuclear so that the
definition of single-particle levels, as in free (laboratory) nuclei,
is not unreasonable.  The form of single-particle wave functions changes
completely, owing to spatial quantization, in the progression to
a new shell and this is reflected in the level density discontinuity
which produces shell effects.  But in reality, single-particle or hole
states are
certainly modified by residual nucleon-nucleon interactions which
introduce components of the same angular momentum and parity but with
more complex particle-hole structure.  Our expectation
is that this configuration mixing would change the
relative spacings of the $\epsilon_{i}$ so as to
reduce the values of $E_{sp} - \tilde{E}_{sp}$ generated by this
procedure.  Consequently, the extent of shell structure seen in, for example,
Fig. 1, should be treated as an upper limit to the true
contribution of shell structure to formation enthalpy
differences between nuclei of neighbouring $Z$.  A further reason why
these results should be regarded as no more than a guide to true formation
enthalpy differences is that their $Z$-average (in Fig. 1) does not
appear to conform well with $(H_{FZ})_{CLDM}$, possibly because the CLDM
parameters are to some extent inconsistent with the Negele \& Vautherin
levels.

\section[]{Weak-interaction transition rates and cooling}

Given the formation enthalpies for nuclei in the interval $20 \leq Z \leq 50$,
an estimate of the initial condition as the star cools can be
obtained by assuming that, above a certain temperature, weak-interaction
transition rates are large enough to maintain approximate local thermal
equilibrium.  We assume here, quite arbitrarily, that this temperature
is $T_{0} = 5 \times 10^{9}$ K.  The arbitrary nature of our assumption
follows from the difficulty in calculating transition rates at
temperatures $T \stackrel{>}{\sim} T_{0}$ where neutrino phase-space
occupation numbers cannot be assumed to be zero.  It is also necessary
to make the simplifying approximation of neglecting nuclear excited states.
Individual nuclear partition functions are then equal to $2J+1$, where the
nuclear spin $J$ is derived entirely from the protons.  Unpaired neutrons,
or neutron quasiparticles at $T < T_{cn}$,
are viewed as excitations of the neutron continuum rather than of
individual nuclei.  The further evolution of the system at $T < T_{0}$
depends on the specific heat, the neutrino emissivities, and on the
set of weak-interaction $Z \rightleftharpoons Z+1$ transition rates.

Equations (A4)-(A9) of Paper I give the $Z \rightarrow Z+1$ transition rate
from an initial proton closed-shell nucleus.  A proton
is created in a new shell of angular momentum $j$ with, in the superfluid case
at $T < T_{cn}$, either the creation or annihilation of a neutron quasiparticle. 
The conservation equations for these two energetically distinct processes are
\begin{eqnarray}
\pm \epsilon_{n} = H_{FZ+1} - H_{FZ} + \epsilon_{e} + \epsilon_{\bar{\nu}},
\end{eqnarray}
in which the electron and neutron energies are measured from their chemical
potentials $\mu_{e}$ and $\mu_{n}$, with $\epsilon_{n} \geq  \Delta_{n}$,
where $\Delta_{n}$ is the neutron energy gap.  These are a form of direct
Urca transition
because the Fourier transform of the proton wave function always has a finite
amplitude at the wavenumber ${\bf p}_{n} - {\bf p}_{e} - {\bf p}_{\nu}$
necessary for momentum conservation.  The transition rate, summed over all
states in the new shell, is the product of a rate constant $\Gamma_{0}$ and a
phase-space integral (equations A7 and A8).  We refer to Paper I for further
details. 
In the more general case of a partially filled shell, the protons are
assumed to be paired into states of zero angular momentum. The rate constant
$\Gamma_{0}$ is then multiplied by a
$j$-dependent factor.  For example, in a $Z \rightarrow Z+1$ transition in
which the shell initially contains two protons, the factor is 
$(1 - 2/(2j+1))$.  A second $Z \rightarrow Z+1$ case is that in which the
shell initially contains an odd number of protons.  For the case of a single
proton, the factor is $2/(2j+1)^{2}$.

Nuclear spins obtained directly from the Negele \& Vautherin shell ordering
are assumed in the present paper, but with some reservation.  The sequences
of ground-state spins and parities tabulated by M\o ller et al (1997) for
nuclei close to neutron instability, and hence relevant below neutron-drip,
are not at all simple, possibly owing to the significant nuclear deformation.
It is not obvious that, above neutron-drip, a system of $Z$ protons embedded
in a neutron continuum will be without deformation and consequent complexities.
But the conclusions which will be described in this Section concerning
$Z$-heterogeneity are so clear
that they would not be affected by changes in the detail of shell ordering.

The parameters used in the evaluation of transition rates at a given
temperature are those listed in Table 1 except that we assume different
neutron energy gaps and critical temperatures in the nuclear interior and
in the continuum.  The neutron effective mass is
$m_{n}^{*} = 0.8m_{n}$.   Following a local density approximation (LDA), values
of the neutron Fermi wavenumber $p_{Fn}$ and
the zero-temperature neutron
energy gap $\Delta_{n}$ are those for an infinite system at the density of
the nuclear interior.  The critical
temperature $T_{cn}$ is that for an isotropic BCS superfluid.  The
finite-temperature energy gap needed for calculation of the phase-space
integral is conveniently obtained from tabulated results given by
Rickayzen (1965), as is the BCS neutron specific heat.  The specific heat
of the system has components for the electron gas, the solid, and for the
neutrons (normal or BCS superfluid).  Although the solid is amorphous and
heterogeneous, it is represented by a simple Debye model with a single
temperature $T_{D}$.

The neutron energy gap in neutron-star matter is not well known.  Different
methods of calculation lead to a wide range of values, (see the reviews of
Pethick \& Ravenhall 1995,
Heiselberg \& Hjorth-Jensen 2000).  However, methods that 
introduce corrections such as medium polarization to the calculation
inevitably seem to suppress the energy gap as shown, for example, by
Shen et al (2003) who also give a brief review of recent work on this topic.
For this reason, we have assumed that the energy gap obtained by Ainsworth,
Wambach \& Pines (1989) applies to the neutron continuum $\Delta_{n}^{e}$.
This affects only
the specific heat.  The neutron energy gap effective in the nuclear
interior and the proton quasiparticle energy are even more difficult to
estimate in neutron star matter.  The neutron coherence length is at least of
the same
order of magnitude as the nuclear radius, so that the proximity effect is
certain to be significant for the neutron gap, as has been emphasized by
Pethick \& Ravenhall.  Hence the value of the LDA assumption must be limited.
These authors also observe that pairing in the interior of ordinary nuclei is
a kind of proximity effect in that attractive pairing in the relatively
large, low-density, surface volume influences pairing in the whole.  The
absence of this low nucleon-density region owing to the presence of the
neutron continuum may then affect the proton pairing $\epsilon_{pq}$.
The uncertainties of prediction are such we think it necessary to
treat the interior neutron energy gap and the proton quasiparticle energy
as unknown parameters which are both assumed to be within the interval 
$0 < \epsilon_{pq},\Delta_{n}< 1$ MeV.

\begin{figure}
\scalebox{0.50}{\includegraphics{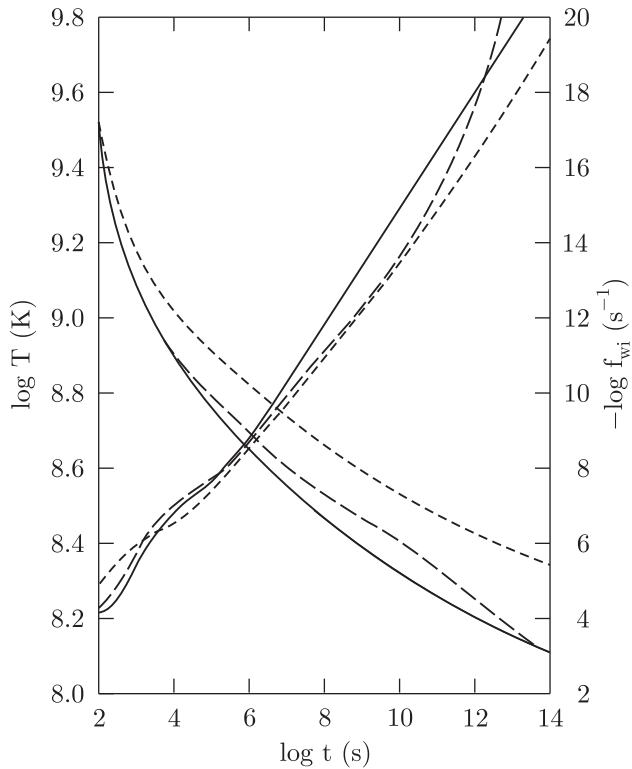}}
\caption{The left-hand set of curves shows the cooling as a function of
time at a matter density of $3.7 \times 10^{13}$ g cm$^{-3}$ for formation
enthalpies given by the solid circles of Fig. 1.  The cooling is adiabatic
except for the neutrino and antineutrino emissivities described by
equation (10).  Neutrino pair production by electron bremsstrahlung and
by neutron quasiparticle annihilation are neglected here.  Allowance for
these processes would produce faster cooling, particularly at
$t\stackrel{>}{\sim} 10^{10}$ s. The right-hand
curves measure the movement of nuclear charge (the flux $f_{wi}$ per unit nucleus)
from $Z = 40$
toward the closed shell at $Z = 34$ as a consequence of the weak interaction.
It is a function of temperature, but is represented in the figure as a function
of time by means of the left-hand set of curves.  The product $tf_{wi}$ is
always some orders of magnitude smaller than unity, showing that the
movement of charge during cooling is insignificant.  The solid, large and
small-dashed curves are, respectively, for the parameter sets:
$\Delta_{n}=0.2$, $ \epsilon_{pq}=0.4$ MeV; $\Delta_{n}=0.4$, $ \epsilon_{pq}
=0.4$ MeV; $\Delta_{n}=0.4$, $ \epsilon_{pq}=0.6$ MeV.}
\end{figure}

Routines for the evaluation, at any temperature, of the specific heat and
of all transition rates for $20 \leq Z \leq 50$ with the proton
level sequence $1{\rm f}_{7/2}, 1{\rm f}_{5/2}, 2{\rm p}_{3/2},
2{\rm p}_{1/2}, 1{\rm g}_{9/2}$ allow the change in $Z$-distribution to
be followed as the star cools.  They also give the neutrino emissivity.
We have adopted an arbitrary fixed initial
temperature $T_{0} = 5 \times 10^{9}$ K, which is close to the {\it bcc} 
lattice melting temperatures given in Table 1. 
The cooling curves for a matter density of $3.7 \times 10^{13}$
g cm$^{-3}$ (adiabatic except for neutrino and antineutrino emission through
the processes described by equation 10) are
shown in Fig. 2. for the set of $H_{FZ}$ values given in Fig.1 which have no
shell correction except for the $\epsilon_{pq}$ unpaired-proton term.
Three sets of the parameters $\epsilon_{pq}$ and $\Delta_{n}$ are considered.
The early parts of these curves are not entirely
reliable because our calculations neglect neutrino opacity and so
overestimate transition rates and emissivities at temperatures
significantly above $10^{9}$ K.  The right-hand scale gives a measure of
how rapidly, at any given temperature, the weak interaction is changing the
nuclear $Z$-distribution by
moving nuclei between two of the minima in $Z$ which appear in Fig. 1.  It
shows the flux $f_{wi}$ between $Z = 40$ and $Z = 34$, calculated at $Z = 37$
and normalized per unit nucleus in the system, for the cooling conditions
assumed here.  This quantity is a function of temperature, apart from a
relatively small dependence on the changing $Z$-distribution.  It is given
here as a function of 
time by using the left-hand sets of curves in the Figure.  Using the
(right-hand) sets of curves so obtained, it can be seen that  
the product $tf_{wi}$ is always some orders
of magnitude smaller than unity.  (Fig. 2 allows recovery of $f_{wi}$
as a function of $T$ so that it is possible to estimate the time required for
weak-interaction equilibrium at a fixed temperature.) Weak-interaction rates 
are so low that even with the extreme assumption of no shell corrections,
the distributions of fractional concentrations for even-$Z$ nuclei
change very little with time provided $\epsilon_{pq}\ge 0.4$ MeV.  In the limit of
large $t$, the resulting values of the impurity parameter $Q$ are not much smaller
than those ($Q = Q_{m}$ in Table 2 of Paper I) for thermal equilibrium at the
melting temperatures without shell corrections and with $\epsilon_{pq} = 0$.
They are shown here in Table 2. In the shell correction cases, there is
naturally some movement of $Z$ toward the magic numbers but the
resulting values of $Q$ are
independent of $\Delta_{n}$ and $\epsilon_{pq}$ to such an extent that, given
the very approximate nature of our calculations, it is not useful to
include in the tabulation the values of these two parameters used, which are
those of Figs. 4-6.  The same comment can be made for the no shell-correction
cases provided, as we stated previously, that $\epsilon_{pq}\ge 0.4$ MeV.
Lower values of this parameter, in the special case of
the absence of shell corrections, produce potential barriers small enough to
allow movement of $Z$ toward $\bar{Z}$.  But we do not attempt to give
$Q$-values for this set of assumptions which are very unlikely to be
physically realized.
The $Q$-values of Table 2 and their significance will be considered
further in Section 6.  But it is not realistic to neglect shell effects
completely.  These must be present although, owing to uncertainties such as
configurational mixing (Section 2), the formation enthalpy differences they
introduce may be smaller than those shown in Fig.1.  Thus the potential
barriers in the weak-interaction paths between the closed shells at
$Z = 20, 28, 34, 40, 50$ must be much higher than assumed in our calculations
of $f_{wi}$, and the actual values of $f_{wi}$ many orders of magnitude smaller
as a consequence.

\begin{table}
\begin{minipage}{80mm}
\caption{Values of the impurity parameter $Q$ and mean charge $\bar{Z}$:
$Q=Q_{sc}$ with shell corrections; $Q=Q_{nsc}$ with no shell corrections.}
\begin{tabular}{@{}lllll@{}}
\hline
$\rho$ & $Q_{sc}$ & $\bar{Z}_{sc}$ & $Q_{nsc}$ & $\bar{Z}_{nsc}$\\
$10^{13}$ g cm$^{-3}$ & & & & \\
\hline
1.6 & 12 & 37.8 & 5 & 34.6 \\
3.7 & 6 & 35.3 & 17 & 33.8 \\
8.8 & 19 & 39.0 & 24 & 34.4 \\
\hline
\end{tabular}
\end{minipage}
\end{table}

Sets of curves similar to those of Fig. 2 exist for the other densities
listed in Table 1.  But they are so similar that they are not shown here.
Our conclusion is that, even
with allowance for the uncertainties inherent in the shell-effect
formation enthalpy differences calculated here, solid neutron-star
matter cools to a system heterogeneous in $Z$.  Other sources of error are
much less significant.  For example, the Bessel function bound-proton
states assumed in equations (A4)-(A6) of Paper I must overestimate the true
transition rates, as would our neglect of neutrino opacity at $T > 10^{9}$ K.
But modest errors are not important because these transition rates determine
both the cooling and change in $Z$-distribution of the system.  Very broadly,
we can see that a sufficient condition for maintaining $Z$-heterogeneity during
cooling is that the thermal energy per average Wigner-Seitz cell should
be smaller than the mean energy, of the order of $6k_{B}T$, removed by
neutrinos or antineutrinos in a weak transition.  The specific
heat per average
Wigner-Seitz cell is shown in Fig. 3 as a function of temperature.
It demonstrates that the thermal energy satisfies this condition easily
at $T\stackrel{<}{\sim} 10^{9}$ K but not at $T \gg 10^{9}$ K (the
neutron contribution is large at $T \stackrel{>}{\sim} T_{cn}^{e}$).  
It is also true that we have neglected other
well-established sources of neutrino emissivity which, if included, would
cool the system more quickly and so assist in maintaining heterogeneity.

\begin{figure}
\scalebox{0.56}{\includegraphics{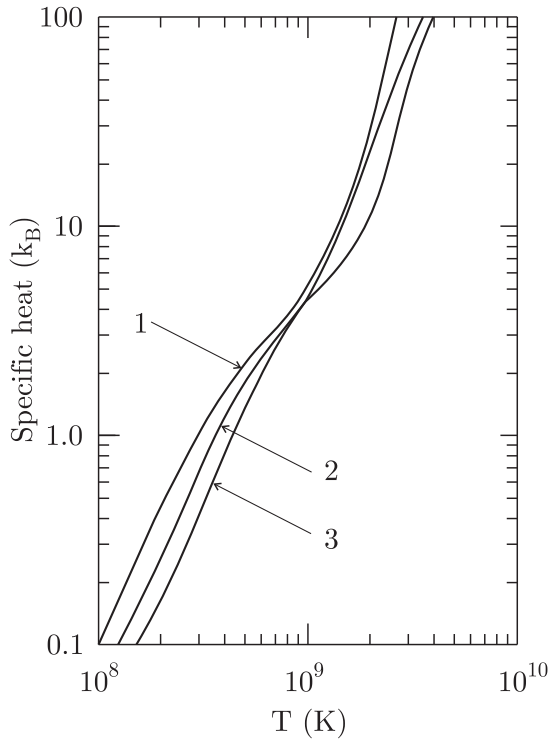}}
\caption{The specific heat per average Wigner-Seitz cell is shown as a
function of temperature, in units of $k_{B}$.  The curves labelled $1-3$
are, respectively, for matter densities $1.6, 3.7$, and $8.8 \times 10^{13}$
g cm$^{-3}$.  The neutron continuum energy gaps $\Delta_{n}^{e}$ are those
given in Table 1.  The internal gap $\Delta_{n} = 0.2$ MeV.}
\end{figure}

\begin{figure}
\scalebox{0.56}{\includegraphics{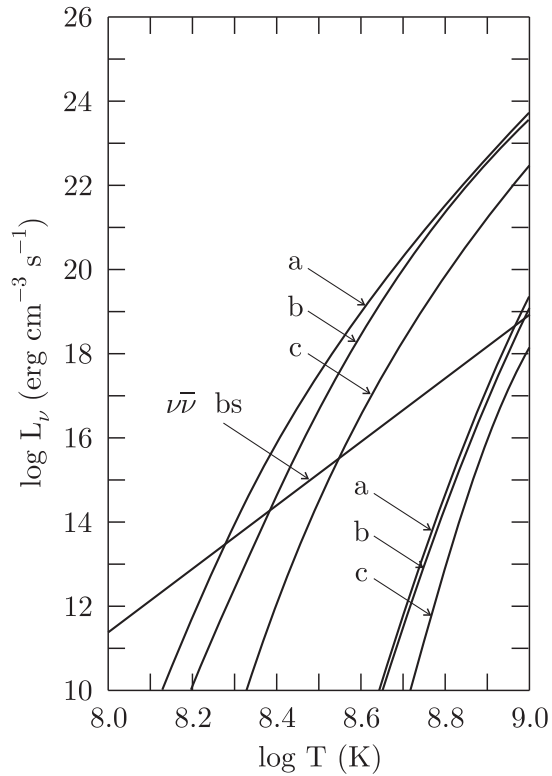}}
\caption{Neutrino emissivities at a matter density of $1.6 \times 10^{13}$
g cm$^{-3}$ are shown as functions of temperature for the weak-interaction
processes described by
equation (10). The formation enthalpies are obtained by the inclusion of
shell corrections (lower set of curves) and without these corrections (upper
set of curves).  Neutrino-pair production by electron bremsstrahlung and
neutron quasiparticle annihilation are
excluded from these curves, but their emissivity is shown separately for
reference purposes.
The curves labelled $a-c$ are, repectively, for the following parameter sets:
$\Delta_{n}=0.2$, $ \epsilon_{pq}=0.4$ MeV; $\Delta_{n}=0.4$, $\epsilon_{pq}=0.4$
MeV; $\Delta_{n}=0.4$, $\epsilon_{pq}=0.6$ MeV. }
\end{figure}

\begin{figure}
\scalebox{0.56}{\includegraphics{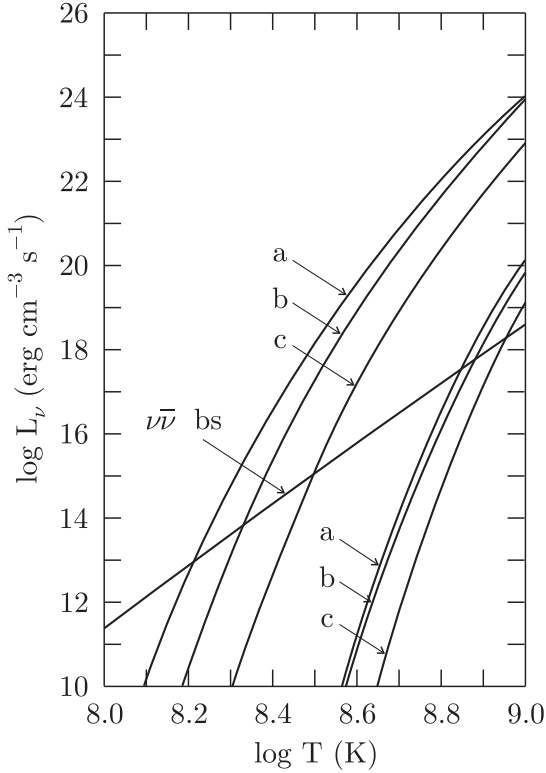}}
\caption{Neutrino emissivities at a matter density of $3.7 \times 10^{13}$
g cm$^{-3}$ are shown with labelling and parameter sets as for Fig. 4.}
\end{figure}

\begin{figure}
\scalebox{0.56}{\includegraphics{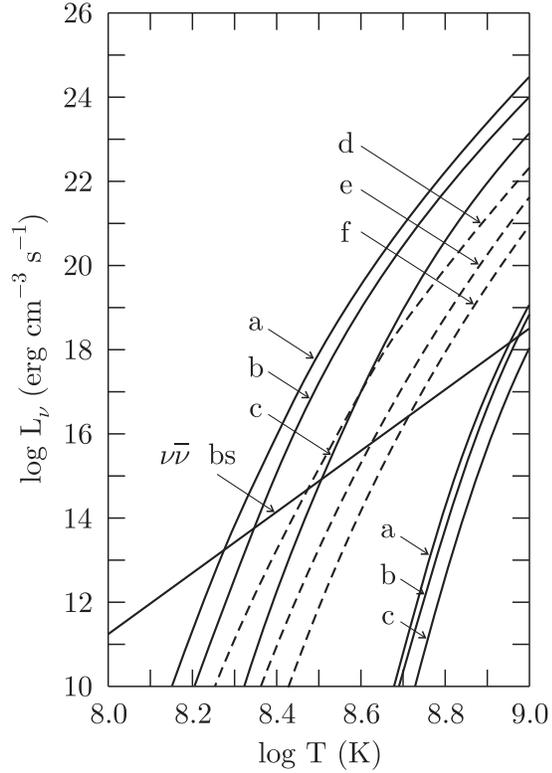}}
\caption{Neutrino emissivities at a matter density of $8.8 \times 10^{13}$
g cm$^{-3}$ are shown with labelling and parameter sets as for Fig. 4.
The additional set of broken curves are the neutrino emissivity $L_{s\nu}$
given by equation (11) for a one-dimensional slab nuclear structure.
The curves labelled $d-f$ are, respectively, for the parameter sets:
$\Delta_{n}=0.2$, $\Delta_{p}=0.4$ MeV; $\Delta_{n}=0.4$, $\Delta_{p}=0.4$
MeV; $\Delta_{n}=0.4$, $\Delta_{p}=0.6$ MeV.}
\end{figure}

The neutrino emissivities are shown in Figs. 4-6 for the matter densities
listed in Table 1.  In each case, they are given both with and
without the Strutinsky shell corrections obtained here and for sets of
values of $\epsilon_{pq}$ and $\Delta_{n}$.  The principal well-established
neutrino-emissivity processes, for the inner crust, in
studies of neutron-star cooling are neutrino-pair production by electron
bremsstrahlung and by neutron quasiparticle annihilation in the superfluid.
We emphasize that these emissivities have not been included in our cooling
calculations but,
for reference purposes, they are shown separately, obtained from
Fig. 4 of the paper by Kaminker, Yakovlev \& Gnedin (2002).  Obviously, the
potential barriers which separate formation enthalpy minima at closed shells
are of crucial importance in determining emissivities.  It is possible to
say with confidence only that
the emissivities shown are likely to form upper and lower bounds for the
true values.  As they differ by at least several orders of magnitude, this
is not a very strong or practically useful statement.  But given that the solid
inner
crust is heterogeneous in $Z$, the existence of this source of uncertainty must
be accepted.  Its effect on the surface temperature of
the star is difficult to estimate without complete cooling calculations.

\section[]{Low-dimensional structures}

Phases in which protons are confined to one or two-dimensional structures
have been studied extensively and their existence in a
substantial density interval between the spherical nuclear phase and the
liquid core depends on the form of the Skyrme pseudopotential
assumed in deriving the equation of state.  Both Lorenz et al (1993),
using the interaction of Lattimer et al, and Oyamatsu (1993) find
such structures within a significant density interval. A survey
of work on this topic has been given by Pethick \& Ravenhall (1995).
But more recent calculations by Douchin \& Haensel (2000),
using a different Skyrme interaction, find only the phase of spherical
nuclei.
Investigations since then (Watanabe \& Iida 2003) have shown that
the inclusion of electron screening tends to increase the density
interval occupied by any such phase.
The paper by Lorenz et al notes that these structures give rise to
new weak-interaction processes but gives no further details.  These
are no more than a different form of the direct Urca transition
considered in Section 3.  For completeness, the neutrino emissivity has been
calculated here for one of these cases, allowing a comparison with the
range of emissivities obtained for the spherical nuclear phase.  The
one-dimensional system considered is that in which the Wigner-Seitz cell
is an infinite slab of thickness $2d_{WS}$ containing neutrons and
relativistic electrons.  Protons are confined within a slab of
thickness $2d_{N}$ symmetrically positioned inside this cell.
Thus the component of the proton wave function in the variable perpendicular
to the slab has Fourier transform $\psi_{m}$ with quantum number $m$ and
finite amplitude at
the wavenumber necessary for momentum conservation.  We assume that, with
this limitation, neutrons and protons each form an isotropic BCS superfluid,
with energy gaps $\Delta_{n,p}$.  The neutrino emissivity per unit volume
is then
\begin{eqnarray}
L_{s\nu} & = & \frac{(1 + 3C_{A}^{2})G^{2}\cos^{2}\theta_{c}}{(2\pi)^{11}
\hbar d_{WS}}
\int d^{3}p_{n}d^{2}qd^{3}p_{e}d^{3}p_{\nu} \nonumber \\
 &   &  \sum_{m}\left|\psi_{m}(p_{n\perp} -
p_{e\perp})\right|^{2} \delta^{(2)}({\bf p}_{n} - {\bf p}_{e} -
{\bf q}) \nonumber \\
 &   &  \sum \epsilon_{\nu}u^{2}_{n}v^{2}_{p}n_{n}(1 - n_{p})(1 - n_{e})
 \nonumber \\
 &   &  \delta(\epsilon_{n}-\epsilon_{p}-\epsilon_{e}-\epsilon_{\nu}).
\end{eqnarray}
In this expression, $G$ is the muon decay constant, $\theta_{c}$ is
the Cabibbo angle, and $C_{A} = 1.25$ is the ratio of axial vector to vector
coupling constants.  The wavevector ${\bf q}$ lies in the plane of the slab.
The neutrino or antineutrino energy is $\epsilon_{\nu}$.  Quasiparticle
or electron occupation numbers are $n_{n,p,e}$ and $\epsilon_{n,p,e}$
are the energies, referred to the chemical potential.  The Bogoliubov
coefficients for the proton and neutron quasiparticle states are $u_{n}$ 
and $v_{p}$.  The terms shown in this expression are for the process
of neutron quasiparticle annihilation with proton quasiparticle and
electron creation.  However, the unlabelled summation states that the set of 8 
terms, derivable from the creation or annihilation of the electron
and of the neutron and proton quasiparticles, are included.

Numerical evaluation of this somewhat untidy expression gives the
emissivities shown in Fig. 6 for a matter density of $1.59 \times 10^{14}$
g cm$^{-3}$ ($d_{N} = 3.62$ fm, $d_{WS} = 10.1$ fm) for
several sets of values of $\Delta_{n,p}$.  Values of the emissivity
do not vary greatly with matter density in this phase or with the
parameter $d_{N}$ which is itself very much dependent on CLDM details.
Emissivity calculations with different values of $d_{N}$ show that uncertainties
arising here are small compared with those caused by our lack of
knowledge of the energy gaps $\Delta_{n,p}$.  In general, the emissivities
are rather smaller than those found in the spherical nuclear phase, the
reason being that
the Fourier transforms of the proton states in that phase have broader
distributions of momentum.

\section[]{The outer crust}

The canonical crust structure at densities $\rho < \rho_{nd}$ is obtained by
minimization of the Gibbs free energy per nucleon.  It consists of
successive spherical shells, each of a {\it bcc} lattice homogeneous in the
nuclear charge $Z$ neutralized by a relativistic electron gas.  The most
recent calculations are those of Haensel \& Pichon (1994) who also review
earlier work.  The zero-temperature structure of the interfaces between these
shells was examined by Jog \& Smith (1982) and found to consist of extremely thin
layers of interpenetrating simple cubic lattices of the two nuclear charges
concerned.  Formation, however, must be considered at the lattice melting
temperature $T_{m}$, or above.  
The reduction in free energy, derived from the melting temperature
entropy of mixing for the two nuclear species, then gives interface layers
of significant thickness in which one nuclear species is
present as an impurity in a lattice formed by the other (De Blasio 2000).

Consideration of the later stage of nuclear formation at finite temperature
gives quite separate grounds for excluding homogeneous lattices (Jones 1988).
At temperatures in the vicinity of $T_{m}$, nuclei are in state of partial
thermal equilibrium with the electrons and with a low-density Boltzmann gas
of neutrons.  (Weak-interaction transition rates, even at $T_{m}$, are too
small to guarantee equilibrium $Z$-values, but the transition rates for the
emission or absorption of neutrons are high and so
maintain strong-interaction equilibrium with
the Boltzmann gas at any density.)  From the definition of the neutron-drip
density $\rho_{nd}$, the neutron chemical potential in this region is
$\mu_{n} = \mu_{n}^{e} < 0$, referred to the rest energy as zero.
The equilibrium number density of the Boltzmann gas is 
\begin{eqnarray}
n_{n}^{e} = \frac{2}{\lambda_{n}^{3}}\exp(\beta\mu_{n}^{e}),
\end{eqnarray}
where
\begin{eqnarray}
\lambda_{n} = \sqrt{\frac{2\pi \beta \hbar^{2}}{m_{n}}},
\end{eqnarray}
and $\beta^{-1} = k_{B}T$.  Thus it decreases rapidly as the star cools.
But the nature of the system imposes an additional constraint on the
Gibbs function minimization.  The neutron scattering cross-section is so
large that, once nuclei have formed, diffusion over macroscopic distances
of the order of the
shell depth is not possible within the short time permitted by neutron
condensation on to nuclei.  This occurs at a temperature $\tilde{T} > T_{0}$,
where
$T_{0}$ is the temperature (not well-defined) at which weak-interaction
equilibrium fails.
Consequently, it is necessary to impose the further
condition, at temperatures $T \stackrel{<}{\sim}T_{m}$, that the mean
nucleon number $A_{WS}$, per Wigner-Seitz cell,
should be constant.  We refer to Jones (1988) for further details.

The method of calculating equilibrium nuclear number densities differs
from that of Paper I which was for densities above $\rho_{nd}$.  The
pressure at $\rho < \rho_{nd}$ is small in
nuclear structure terms, being almost
entirely that of the degenerate electron gas.  Nuclei in this region
have definite $A,Z$ and
can be assumed to have binding energies $B(A,Z)$ identical with those of
the free
(terrestrial) state.  
The minimization of the Gibbs function $G$ for a fixed number of baryons is
at constant $\mu_{e}$ and hence approximately at constant pressure.  It is,
\begin{eqnarray}
\frac{\partial}{\partial n_{ij}}\left(G+\Lambda\sum_{ij}n_{ij}\right) = 0,
\end{eqnarray}
where $n_{ij}$ is the number of nuclei with mass number $A_{i}$ and charge
$Z_{j}$, and $\Lambda$ is a Lagrange multiplier.  The equilibrium
condition for the nuclear chemical potential $\tilde{\mu}_{ij}$ is,
\begin{eqnarray}
\tilde{\mu}_{ij}+(A_{WS}-A_{i})(\mu_{n}^{e}+k_{B}T)+Z_{j}\mu_{e}+\Lambda = 0.
\end{eqnarray}
(Without the constant $A_{WS}$ constraint, the nuclear chemical potential 
would be $\tilde{\mu}_{ij} = A_{i}\mu_{n}^{e} - Z_{j}\mu_{e}$.)
The neutron chemical potential is not an independent variable
in equation (15): it is
a function of $A_{WS}$ through the relation $A_{WS}-\bar{A} = n_{n}^{e}V_{WS}$,
in which $\bar{A}$ is the mean nuclear mass number and $V_{WS}$ the
Wigner-Seitz cell volume.
To a satisfactory approximation, the nuclei are independent systems and
their equilibrium number densities are given by an
expression entirely analogous with equation (12),
\begin{eqnarray}
n_{A,Z} = \frac{\mathcal{Z}}{\lambda_{A,Z}^{3}}\exp
\left(\beta(\tilde{\mu}_{ij} + Z(m_{n} - m_{p}) + B - E_{WS})\right),
\end{eqnarray}
where $\mathcal{Z}$ is the nuclear partition function (normalized so
that at $T = 0$ it is $2J+1$), $B$ is the
nuclear ground-state binding energy, and $E_{WS}$ is the
Coulomb energy of the Wigner-Seitz unit cell.  The presence of $\Lambda$
allows two finite number densities in the  zero-temperature limit, as
is necessary because $A_{WS}$ is in general not an integer. 
Evaluation of equation (16) at a standard temperature $T_{0}$ ,
close to $T_{m}$, for even-even nuclei with $\mathcal{Z}=1$ and binding
energies from the
recent compilation of M\o ller et al (1997) gives estimates of
the mass number and charge heterogeneity expected at densities below
$\rho_{nd}$.

It is unfortunate that the results are inconclusive.  The source of the
problem is that the values of $A_{WS}$ are determined at those temperatures
$\tilde{T} > T_{0}$ existing at the formation of nuclei, which are high
and poorly known.  The bulk transport of neutrons over macroscopic distances,
of the order of shell depths, is not possible in the short time which elapses
before rapid cooling produces almost complete condensation through nuclear
capture (Jones 1988).  It has to be accepted that the values of $A_{WS}$,
constant at $T < \tilde{T}$, are virtually unknown.  Evaluations of equation (16)
for matter densities $\rho \sim 1-3 \times 10^{11}$ g cm$^{-3}$, temperatures
$T_{0} = 5 \times 10^{9}$ and $10^{10}$ K, and a wide range of values of
$A_{WS} - \bar{A}$ give a common picture.  As $A_{WS} - \bar{A}$ increases,
the equilibrium nuclei change from a group associated with the neutron $N=50$
closed shell to a group near $N=82$.  The neutron chemical potentials at
changeover vary from $-1.3$ MeV at $3\times 10^{11}$ g cm$^{-3}$ to $-3.6$
MeV at $1 \times 10^{11}$ g cm$^{-3}$, for $T_{0} = 5\times 10^{9}$ K.
The corresponding values at $10^{10}$ K are $-1.8$ and $-4.1$ Mev,
respectively.  The changeover from $N=50$ to $N=82$ is similar to that
found by Haensel \& Pichon (1994) in their study of zero-temperature
equilibrium.  The problem is that the $A_{WS}$ constraint which
was not considered by them does not permit us to estimate where it occurs.
This complexity is unfortunate, but we believe that it is real.
Qualitatively, the consequences for $Q$-values are as follows.  The $N=50$
region has $Q \ll 1$, and the $N=82$ region, $Q \sim 1$.  Values
$Q \gg 1$ exist in the changeover region owing to the very large
differences in $Z$ which are present there, as found earlier by
De Blasio (2000).  Detailed calculations
of weak-interaction transition
rates confirming the absence, at $T < T_{0}$, of weak-interaction
equilibrium have not been made for $\rho < \rho_{nd}$, but the qualitative
criterion given in Section 3 is satisfied.  In particular, values
$Q \gg 1$, where they exist, will certainly remain frozen in.

\section[]{Transport coefficients and conclusions}

The main conclusion of this paper is that nuclear-charge heterogeneity
exists in the solid phase of isolated neutron stars.
There must be many reservations
about the procedures described in Sections 2 \& 3 which have been used to
obtain this result.  We attempt to summarize them here and in each case
give reasons why they should not be viewed as serious.  Firstly, the
shell-corrected formation enthalpy values, $H_{FZ}$, are based on the
Negele \& Vautherin shell ordering and spacings.  Nuclei close to
neutron-instability are known to be deformed, with ground states of
some complexity (see M\o ller et al 1997).  It is quite possible that
these features are also present above $\rho_{nd}$.  Therefore, the values
shown in Fig. 1 may not give a true picture of the potential barriers
which slow weak-interaction transitions.  But it is the case that even quite
small barriers slow weak transitions sufficiently.  The transition rate
calculations, following the procedure of Paper I, are based on elementary
single-particle proton wave functions and neglect Coulomb corrections to
the electron function, which are only moderate at the values of $\mu_{e}$
considered.  But here, the effect of more complex ground states
would inevitably be reductions in transition rates which would be
largely neutral in effect because the rate of cooling by neutrino
emission is also reduced.  The results, of which those in Fig. 2
are an example, show that the failure of weak-interaction equilibrium
during cooling at $T < T_{0}$ is so clear that the above uncertainties are not
significant.  The choice of $T_{0} = 5 \times 10^{9}$ K as the lowest temperature
at which complete equilibrium remains is, of course, arbitrary.  There is no
doubt that a higher value would give more heterogeneity in $Z$,
with larger values of $Q$, and that calculations made by the present methods
would show a failure of weak-interaction equilibrium
at temperatures below it.  But at these very high temperatures, the neutrino
opacity would cease to be negligible, as assumed in Section 3, so that
the cooling rates shown in Fig. 2 would be over-estimates.  Obviously,
the $Q$-values given in Table 2, under different stated assumptions, for
each of the matter densities listed in Table 1, should be seen as little better
than order of magnitude estimates.  But these satisfy $Q \gg 1$ in a way which
does not depend systematically on the details of shell effects assumed.
They are some orders of magnitude larger than the values
$Q \ll 1$ previously assumed in the standard view of solid neutron-star
matter. Although the relevant calculations have not been made,
we suggest there is no reason to think that this conclusion would be
changed by replacing the Lattimer et al CLDM with that of Douchin \&
Haensel (2001).  In the latter case, the important proton closed shells
would be those at $Z=40, 50$ but we anticipate that the orders of
magnitude of the $Q$-parameter obtained would be the same.

The procedures of Sections 2 \& 3 are concerned merely with establishing
$Z$-heterogeneity.  The small-scale structure of the solid is a different
question.  It is not obvious that diffusion at temperatures in the glass
transition region, which we assume to be near the homogeneous lattice
melting temperature $T_{m}$, could produce localized order extending over
linear dimensions much greater than $10^{1-2}$ inter-nuclear separations.
The results of
rudimentary formation enthalpy calculations given in Paper I do not indicate
that large chemical potential gradients exist to drive such diffusion.
Moreover, the classical entropy of disorder remains large.  For these
reasons, it is believed that although the structure may not be exactly that
of an amorphous heterogeneous solid, any local order will be at most of small
linear dimension, analogous with nanostructures in ordinary amorphous matter.

These conclusions can be compared with those of recent work on neutron
stars in binary systems with high mass transfer rates,
$\stackrel{>}{\sim}10^{-8}\rm{M}_{\odot}$ yr$^{-1}$ (Schatz et al 1999,
see also Brown \& Bildsten 1998).  Stable burning of hydrogen and helium
occurs near the surface, the most important process being rapid proton
capture, and is almost complete at a depth of the order of $10^{8}$ g
cm$^{-3}$ (see Schatz et al, Fig.1).  The end products have charge
distributions which depend on the mass-transfer rate, but are typically
broad, with $Z\leq 40$.  Continued mass-accretion forces these nuclei
to higher matter densities, eventually to $\rho > \rho_{nd}$.  Schatz et
al contrast this $Z$-heterogeneous matter with the solid phase of an
isolated (primordial) neutron star having the properties described by
Pethick \& Ravenhall (1995).  Our conclusion is that the solid phases
of neutron stars in these different environments are broadly similar, with
impurity parameters $Q \stackrel{>}{\sim} 10$.  It is very unlikely that
the temperatures of the burning processes will be high enough
to produce weak-interaction equilibrium at matter densities above
$\rho_{nd}$.  The flux $f_{wi}$, representing the movement of nuclear charge
toward a closed proton shell, is shown in Fig. 2 for the case of no shell
correction, apart from the proton pairing term $\epsilon_{pq}$.  This extreme
case is not realistic because there is almost certainly some shell correction,
though perhaps not so large as that shown in Fig. 1.  The increased potential
barriers introduced by the correction would reduce values
of $f_{wi}$ by many orders of magnitude, to an extent that $Z$-heterogeneity
would be largely unaffected. 

It is obvious that the amorphous and heterogeneous nature of the solid phase
of neutron-star matter has significant consequences for its mechanical
properties and, above $\rho_{nd}$, for its interaction with superfluid neutron
vortices.  Mechanical properties will not be considered here except to note
that the response to stress is not strictly that of an ordinary amorphous
solid because neutron-star matter is very far from being absolutely stable
and therefore cannot exhibit brittle fracture (Jones 2003).  Superfluid
neutron vortices interact with nuclei through elementary pinning forces
whose calculation is associated with unresolved problems (for a brief partial
review, see Jones
2002).  It is currently not possible to say that the signs or orders of
magnitude are known with any degree of certainty at a given matter density.
Both vortex pinning and the
dissipative force acting on a moving vortex are strongly influenced by the
structure of the solid phase.  Apart from the kind of low-dimensional
structure considered in Section 4, very small dissipative forces would be
possible only for motion through a system of large single crystals
with very low concentrations of dislocations and point-defects.  The
confirmation of the amorphous and heterogeneous nature of the solid phase
rules out this possibility.

The neutrino emissivity from nuclear weak interactions at matter densities
below $\rho_{nd}$ has not been calculated owing to the difficulty in
knowing the correct transition rates between nuclei which are close to
being neutron-unstable.  The assumption of superallowed Fermi transitions
gives an order of magnitude of $10^{22}$ erg cm$^{-3}$ s$^{-1}$ at $10^{9}$ K  
for matter densities $1-3\times 10^{11}$ g cm$^{-3}$, but allowance for the
forbidden nature of the transitions would probably reduce this to an emissivity
not much different from that for neutrino-pair production by electron
bremsstrahlung under these
conditions ($10^{18}$ erg cm$^{-3}$ s$^{-1}$; see, for example, Gnedin et al 2001).
Neutrino emissivity above $\rho_{nd}$ is shown in Figs. 4-6.  Weak-interaction
transitions in this region involve the creation or annihilation of quasiparticles
in the neutron continuum.  Rate calculations are more straightforward in this
case,
but owing to the uncertainties in formation enthalpy differences considered
above and in Section 2,  the emissivities shown in the two sets of curves
probably form upper
and lower bounds for the true values.  However, at temperatures
$T\stackrel{>}{\sim} 4\times 10^{8}$ K, they can be larger than
those previously assumed.  The degree of uncertainty here is unfortunate
but it does seem right that its existence should be recognized in
neutron star cooling calculations.  Neutrino emissivities for one-dimensional
layer structures (Section 4) are also subject to some uncertainty, but
are rather smaller than for the spherical nucleus phase considered in
Section 3.  Emissivities for a two-dimensional phase have not been
calculated, but are probably of a similar order of magnitude to those for
one dimension.

Electron scattering in a $Z$-heterogeneous system contributes electrical
and thermal resistivities which are related by the Wiedemann-Franz law and
are proportional to $Q$.  The corresponding electrical conductivity
(for zero magnetic flux density) is, for relativistic electrons at
$\rho \stackrel{>}{\sim} 10^{7}$ g cm$^{-3}$,
\begin{eqnarray}
\sigma_{i} = \frac{\bar{Z}c\mu_{e}}{4\pi e^{2}Q\Lambda_{imp}},
\end{eqnarray}
in which the parameter $\Lambda_{imp} \approx 2.03$ in the ultra-relativistic
case (see Urpin \& Yakovlev 1980, Itoh \& Kohyama 1993) and $\bar{Z}$ is the
mean nuclear charge.  For the values of $Q$ obtained here, the relaxation time
underlying equation (17) is so short that its product with the electron cyclotron
angular frequency exceeds unity only for magnetic fields $B \stackrel{>}{\sim}
10^{13}$ G.  Therefore, there is no need to distinguish between longitudinal
and transverse components of the conductivity tensor, except at $B \gg 10^{13
}$ G.  The conductivity is also classical rather than quantum in nature
at densities of the order of $\rho_{nd}$ or above (see Potekhin 1999).
The electron thermal conductivity $\kappa_{i}$ is given, in
terms of $\sigma_{i}$,
by the Wiedemann-Franz law.  These transport coefficients are of importance,
in most cases, at very different times during neutron-star cooling.  The
thermal conductivity $\kappa$ is usually of interest at early times before
the interior of the star, defined for this purpose as matter densities
$\rho \stackrel {>}{\sim} 10^{7}$ g cm$^{-3}$,  becomes approximately isothermal.
The $Q$-dependent resistivity obtained here decreases $\kappa$, but the
significance of the change is not immediately obvious and would require
complete neutron-star cooling calculations for its investigation. At later
times, internal temperature gradients are very small and it is possible that
the consequent changes in them would be relatively unimportant.

The electrical conductivity $\sigma$ is obtained by combining $\sigma_{i}$
with the phonon-scattering conductivity $\sigma_{ph}$.  The assumption of
a homogeneous {\it bcc} lattice with a low point-defect density gives
$\sigma \approx \sigma_{ph}$.   Useful summaries of the properties of
$\sigma_{ph}$ for such a system has been given by Urpin \& Muslimov (1992)
and by Pethick \& Sarling (1995).  Umklapp processes are by far the more
important contribution to resistivity but become energetically disallowed
at temperatures $T \stackrel{<}{\sim} T_{U}$, where
\begin{eqnarray}
T_{U} = 2.2 \times 10^{8}\rho^{1/2}_{14}\left(\frac{Z}{60}\right)^{1/3}
\left(\frac{10Z}{A}\right)  {\rm K}
\end{eqnarray}
and $\rho_{14}$ is the matter density in units of $10^{14}$ g cm$^{-3}$.
At temperatures $T_{U} < T \ll T_{D}$, where $T_{D}$ is the Debye
temperature,
\begin{eqnarray}
\sigma_{ph} = 5.5\times 10^{23}\rho^{7/6}_{14}\left(\frac{10Z}{A}\right)
^{5/3}T^{-2}_{9}   {\rm s}^{-1},
\end{eqnarray}
whereas in the limit $T \ll T_{U}$,
\begin{eqnarray}
\sigma_{ph} = 2.1 \times 10^{28}\rho^{8/3}_{14}\left(\frac{10Z}{A}\right)
^{14/3}T^{-5}_{9}   {\rm s}^{-1}.
\end{eqnarray}
In the neutron-drip region $\rho > \rho_{nd}$, the best value of the mass
number $A$, on which the phonon spectrum depends, is not at all obvious.
Pethick \& Sarling suggest $A = A_{WS}$ but give no reasons why this
choice should be suitable for superfluid neutrons at $T < T_{cn}^{e}$.
This factor introduces considerable uncertainty in the form of the
function $\sigma_{ph}(T)$.  It is also true that the modified phonon spectrum
of an amorphous solid
must produce substantial changes from equations (18) to (20).  These have not
been investigated here because, for the amorphous and $Z$-heterogenous
solid, the impurity resistivity is large so that $\sigma_{i} < \sigma_{ph}$
except at temperatures $T \stackrel{>}{\sim} 10^{9}$ K.
For comparison with equations (19) and (20), evaluation of equation (17)
gives a conductivity
\begin{eqnarray}
\sigma_{i} = 2.0 \times 10^{24}\left(\frac{\bar{Z}\rho_{14}}{A_{WS}}\right)
^{1/3}\left(\frac{\bar{Z}}{Q}\right)  {\rm s}^{-1}.
\end{eqnarray}
The values of $Q$ given in Table 2 lead to high impurity resistivity
which is only weakly-dependent on $A_{WS}$. The fact that it is also
temperature-independent removes a substantial degree of uncertainty from
any model calculations made with it.

Thus the (zero-field) electrical conductivity in most of the inner-crust
volume is $\sigma \approx 10^{24}$ s$^{-1}$ at all times.  This is
particularly relevant to studies of the evolution of neutron-star magnetic
fields which include both the Hall effect, with the possibility of a cascade
to high wavenumber field components, and ohmic dissipation (see Hollerbach
\& R\"{u}diger 2002 and Geppert \& Rheinhardt 2002, who also review earlier
work on this problem).  The Hall
parameter considered by Hollerbach \& R\"{u}diger is
\begin{eqnarray}
\frac{\sigma B}{n_{e}ec} = 2.3 \times 10^{-3}B_{12}\left(\frac{A_{WS}}
{\bar{Z}\rho_{14}}\right)^{2/3}\left(\frac{\bar{Z}}{Q}\right),
\end{eqnarray}
where $n_{e}$ is the electron density and $B_{12}$ the magnetic flux density
in units of $10^{12}$ G.  It exceeds unity only for $B \stackrel{>}{\sim}
10^{14}$ G.  The impurity resistivity found here is also relevant for any
process in which the field distribution in the solid crust develops
high-multipole components by mass-accretion
or by flux expulsion from a superconducting neutron-star
interior, or for models in which it is confined to the crust (see Page, Geppert
\& Zannias 2000, who also give an account of previous work on distributions of
this kind).  Field components of wavelength $2h$ decay ohmically with an
exponential time-constant
\begin{eqnarray}
t_{c} = \frac{4\sigma h^{2}}{\pi c^{2}}
\end{eqnarray}
which, for $h = 10^{5}$ cm, is only $4.5 \times 10^{5}$ yr, a time small
compared with those usually considered in relation to decay of neutron-star
internal or surface fields.  Its temperature-independence would also simplify
evolutionary calculations such as those of Page et al.  For the
reasons summarized here, we suggest that the structure of the solid phase
of neutron-star
matter is not an irrelevant detail.  The standard assumption of a homogeneous
{\it bcc} lattice is not adequate and the consequences of an amorphous and
$Z$-heterogeneous structure should be included in very many different studies
of neutron-star physics.

\section*{Acknowledgments}
It is a pleasure to thank Dr J Rikovska Stone for introducing me to the
Strutinsky procedure.

\bsp

\label{lastpage}

\end{document}